\documentstyle[aps,prb,psfig,epsfig,amsmath,amssymb]{revtex}
\def\HMG{H_{MG}}
\def\S{{\bf S}}
\begin{document}
\title{Quantum spin models with exact dimer ground states}
\author{Brijesh Kumar\footnote{\tt brij@physics.iisc.ernet.in}\\ Department of Physics\\
 Indian Institute of Science, Bangalore-560012, India}
\date{\today}
\draft
\maketitle
\begin{abstract}
Inspired by the exact solution of the Majumdar-Ghosh model, a family of one-dimensional, translationally
invariant spin hamiltonians is constructed. The exchange coupling in these models is antiferromagnetic, and 
decreases linearly with the separation between the spins. The coupling becomes identically zero beyond a certain distance. 
It is rigorously proved that the dimer configuration is an exact, superstable
 ground state configuration of all the members of the family on a periodic chain. The ground state is 
two-fold degenerate, and there exists an energy gap above the ground state. 
The Majumdar-Ghosh hamiltonian with two-fold degenerate dimer ground state is just the first member of the
family.
 The scheme of construction is generalized to two and three dimensions, and illustrated with the help of some
concrete examples. The first member in two dimensions is the Shastry-Sutherland model.
Many of these models have exponentially degenerate, exact dimer ground states.
\end{abstract}
\section{Introduction}
The studies of quantum spin models are of great current interest. These studies help us in getting some
 understanding of the magnetic properties of the real, physical systems. Studies of the 
magnetic properties described by the dimer (or valence bond) configurations has been a subject of 
continuing research activities, and is of particular interest recently. A recent example is 
the Shastry-Sutherland\cite{shastry_sutherland,miyahara}
 type models used in understanding the physical properties of SrCu$_2$(BO$_3$)$_2$\cite{kageyama}.
This system is a magnetic insulator with dimer ground state, which is topologically equivalent to that of 
the Shastry-Sutherland (SS) model. Then there are studies related to the Kagom\'e
antiferromagnet\cite{elser,mila}, where magnetic excitations are gapped, and this gap is filled with a large
number of low-lying singlet excitations whose number grows as exponential in the number of sites.
 It is believed that the low-energy
physics of the Heisenberg antiferromagnet on Kagom\'e lattice is of the resonating valence bond (RVB) type.
The quantum dimer models have been applied to study antiferromagnets on triangular lattice
\cite{sondhi}. Again, the idea employed is that of resonating dimer 
 coverings of the lattice. The idea of doping RVB ground state to achieve superconductivity has been a subject of
great consideration in the context of high-T$_c$ superconductors\cite{anderson,kivelson}. 
Though the low temperature behaviour of the undoped high-T$_c$ materials does not show up 
RVB like magnetic properties, the idea is still interesting, and
motivates the search for a doped RVB superconductor\cite{bss_bk}.
 All these studies clearly show the importance of understanding the physics governed by valence bond configurations.
 It makes the search for, and the studies of models with dimer ground state particularly desirable. 

\begin{figure}
 \centerline{\epsfig{file=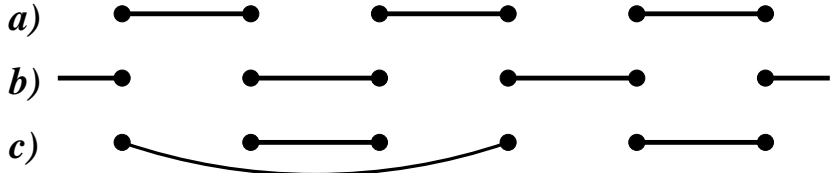,width=11cm}}
 \caption{({\em a}) and ({\em b}) are two exact ground state configurations of the family of 1D models with linear
          exchange coupling, as discussed in the text. 
          The solid line joining two lattice points represents a singlet state between spins
          sitting at corresponding sites. These ground state configurations are being called as 
          $\left|\psi_1\right>$ and $\left|\psi_2\right>$ in the text. 
          ({\em c}) is one of the singlet configurations which is not
          an eigen configuration of the hamiltonian given by Eq.(\ref{hnu}).}
 \label{dimers}
\end{figure}

The Majumdar-Ghosh (MG) model\cite{majumdar_ghosh,majumdar} is a one dimensional quantum spin model with the 
nearest and next nearest neighbour exchange interactions.
 It is exactly solvable for a particular ratio of these exchange couplings,
 and has a two-fold degenerate dimer ground state.
 Though this model has been studied for anisotropic exchange and general spin, $S$, we will consider
 only isotropic exchange and $S=1/2$ case. The exact solution of the MG model guides us in constructing 
models with exact dimer ground states which is the subject of the present work. The hamiltonian for the MG
model is written as :
\begin{equation}
\HMG = J \sum_{i=1}^L \left( 2 \S_i\cdot\S_{i+1} +\alpha\S_i\cdot\S_{i+2} \right)
\label{mgmodel}
\end{equation}
where $J>0$, and $L$ is the number of sites in a one dimensional (1D) lattice with periodic boundary condition
(PBC). It is a well studied model, and shows quantum phase transition from ordered phase to disordered,
 spin-liquid like phase as $\alpha$ is increased from zero to some value greater than 0.482\cite{okamoto}.
At $\alpha=1$, and for $L$ being even, the {\em bond-singlet} (dimer) configurations, as shown in
Fig.(\ref{dimers}-{\em a}, \ref{dimers}-{\em b}), form an exact, two-fold degenerate ground subspace. 
Let us call these dimer configurations as $\left|\psi_1\right>$ and $\left|\psi_2\right>$ which are given below.
\begin{eqnarray}
\left|\psi_1\right> &=& [1,2]\otimes[3,4]\otimes[5,6]\otimes\cdots\otimes[L-1,L] \label{psi1}\\
\left|\psi_2\right> &=& [2,3]\otimes[4,5]\otimes[6,7]\otimes\cdots\otimes[L,1] \label{psi2}
\end{eqnarray}
where \([l,m] = (\;\left|\uparrow_l\downarrow_m\right> - \left|\downarrow_l\uparrow_m\right>\;)/\sqrt{2}\)
is the singlet state of a pair of spins, sitting at sites $l$ and $m$, representing a double bond in the chemical sense. 
The ground state energy, in units of the nearest neighbour exchange (2$J$), is -$(3/8)L$.

 In the following sections, we will discuss, briefly, why $\HMG$ is exactly solvable for its ground state at 
$\alpha=1$. It will guide us in constructing the general family of spin models on 1D lattice with PBC. It will 
be rigorously proved that the dimer configurations, \(\left|\psi_1\right> \) and \(\left|\psi_2\right> \),
 form an exact, two-fold degenerate ground subspace for the whole family. 
Various features of this family of 1D models will be discussed in some detail.
 Then, the generalization to two and three spatial dimensions will be discussed. Our scheme allows us to
construct higher dimensional models with exact knowledge of the ground state. The exponential degeneracy in the
 ground subspace of such models will be discussed. Finally, we will conclude with some general discussion and 
remarks.
\section{The lessons from the MG hamiltonian's solution}
On a chain with PBC, and for $\alpha=1$, the MG hamiltonian can be re-written as:
\( \HMG = J \sum_{i=1}^L\) \( \left( \S_i\cdot\S_{i+1} + \S_i\cdot\S_{i+2} + \S_{i+1}\cdot\S_{i+2} \right)
     = J \sum_{i=1}^L h_i 
\).
The hamiltonian, $h_i = \S_i\cdot\S_{i+1} + \S_i\cdot\S_{i+2} + \S_{i+1}\cdot\S_{i+2}$, is that of a block of three
spins, the $i^{th}$ spin and its next two neighbours, coupled to each other identically. Let us refer to these
 blocks as ${\mathfrak B}_3$ where ${\mathfrak B}$ stands for a block of completely connected spins, and the
subscript, 3, refers to the number of spins in the block. Spins within a block are understood to be
 identically coupled, unless specified. 
 The minimum eigenvalue of $h_i$, $e_i^{min} = -3/4$, for $S=1/2$.
 This is easy to see if we re-write $h_i$ as 
$\frac{1}{2} \left[ \left(\S_i + \S_{i+1} + \S_{i+2}\right)^2 - \S_i^2 - \S_{i+1}^2 - \S_{i+2}^2 \right]$.
The minimum energy spin configuration for ${\mathfrak B}_3$ has one free spin and the rest two spins forming 
a singlet. For example, $ [i,i+1]\otimes\left|\uparrow_{i+2}\right> $ 
is one such eigen-configuration. There are two linearly independent ways of forming such configurations 
for which the total ${\mathfrak B}_3$ spin is 1/2.

Since the MG chain is made up of ${\mathfrak B}_3$ units, the ground configuration of MG chain can be 
constructed in such a way that it is also the lowest energy eigen-configuration of ${\mathfrak B}_3$.
 This is not possible in general. Interestingly, this is possible for MG chain because 
 the minimum energy configuration of ${\mathfrak B}_3$ has strictly one bond-singlet and a {\em free} spin.
This ``free" spin is free in the sense that it can bond with the ``outside" world, and the new composite
 configuration is still the eigen configuration of the block-hamiltonian with energy, $-3/4$, provided other two
 spins of the block form a singlet. Since every spin on the MG chain has identical exchange connectivity, the 
above considerations imply that the ground state configuration of $\HMG$ is the one where every three neighbouring
 spins share exactly one bond-singlet.
 The key observation to make is the fact that {\em the fundamental block has odd number of spins}, and {\em its minimum
energy configuration contains exactly one free spin while rest forming a singlet}. All this straight forwardly 
leads to the exact solution of the ground state of MG chain, already mentioned in the introduction,
 and can be proved rigorously using the inequality, 
 $\left<\phi\right|\HMG\left|\phi\right> \geq E_g \geq J\sum_{i=1}^L e_i^{min}({\mathfrak B}_3) = -(3J/4)L$
 and the identities, 
\( \S_k\cdot(\S_l + \S_m)[l,m]=0\; \forall\; k \neq l\;
\mbox{and}\; m\) ;$\;$ \(\S_l\cdot\S_m[l,m] =
-S(S+1)[l,m]\). If $\left|\phi\right>$ is such an eigen state of the hamiltonian that the upper bound of the
inequality equals the lower bound, then $\left|\phi\right>$ is also the ground state of the hamiltonian.
 In other words, this inequality ensures that {\em an eigen state of the hamiltonian which is also the
 ground state of the basic building blocks, is the ground state of the hamiltonian}.
In the next section, we will construct a general family of one-dimensional spin hamiltonians, and show that
$\left|\psi_1\right>$ and $\left|\psi_2\right>$ are the exact ground state configurations for the whole family.
\section{The family of 1D models with dimer ground state}
\begin{figure}
 \centerline{\epsfig{file=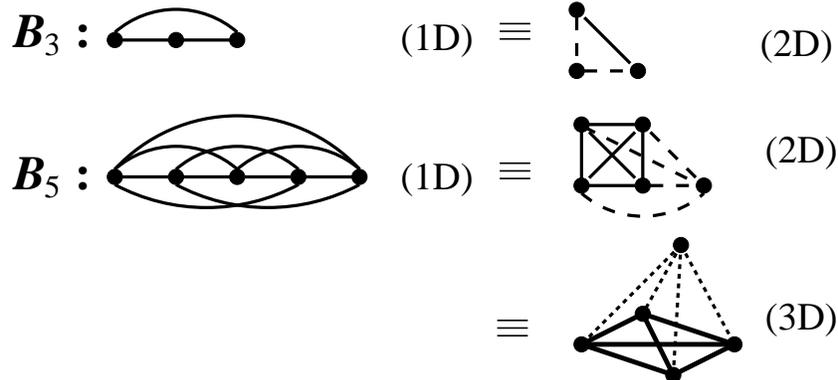,width=11cm}}
 \caption{The ${\mathfrak B}_{2\nu+1}$ blocks for $\nu =1$ and 2, in 1D, 2D and 3D. The dashed coupling in 2D
          blocks is unity while solid one is $\alpha$. In 3D block, coupling of the {\em apex} to the square {\em
          face} is unity while coupling within the face is $\alpha$.} 
 \label{B3_B5}
\end{figure}

{\bf The model :}
Let us consider a block, ${\mathfrak B}_{2\nu+1}$, of identically and completely connected $(2\nu+1)$ spins
where $\nu$ is a positive integer. The identical and complete connectedness of spins means that every spin in a
 block is coupled to every other, with same strength (which is taken to be unity) as shown in 
Fig.(\ref{B3_B5}) for $\nu =1$ and 2. 
On a spin chain, any site $i$ and its
next $2\nu$ neighbours will form a ${\mathfrak B}_{2\nu+1}$ unit once they are identically and completely coupled.
The hamiltonian corresponding to $i^{th}$ such block on a 1D lattice is written as : 
\( h_i({\mathfrak B}_{2\nu+1}) = \left\{\;\S_i\cdot\left(\S_{i+1} +\cdots + \S_{i+2\nu}\right) +\right.\)
                           \( \S_{i+1}\cdot\left(\S_{i+2} + \cdots + \S_{i+2\nu}\right) + \)
                           \( \cdots + \S_{i+2\nu-2}\cdot\left(\S_{i+2\nu-1} + \S_{i+2\nu}\right) + \) 
                           \( \left. \S_{i+2\nu-1}\cdot\S_{i+2\nu}\;\right\} \).
Adding all such block hamiltonians gives the total hamiltonian for the spin chain made up of 
 ${\mathfrak B}_{2\nu+1}$ units.
There are exactly $2\nu$ number of first neighbour (nearest neighbour) pairs, $(2\nu-1)$ number of second
neighbour pairs and so on, within each ${\mathfrak B}_{2\nu+1}$ unit. 
Therefore, the total hamiltonian for a ${\mathfrak B}_{2\nu+1}$ spin chain with PBC is :
\begin{eqnarray}
H[{\mathfrak B}_{2\nu+1}] &=& J\sum_{i=1}^L \left[\;2\nu\S_i\cdot\S_{i+1} + (2\nu-1)\S_i\cdot\S_{i+2}
                             + \cdots + 2\S_i\cdot\S_{i+2\nu-1} + \S_i\cdot\S_{i+2\nu}\;\right] \nonumber\\
                         &=& J\sum_{i=1}^L\sum_{j=1}^{2\nu}(2\nu+1-j)\S_i\cdot\S_{i+j}
\label{hnu}
\end{eqnarray}
Thus, on a ${\mathfrak B}_{2\nu+1}$ spin chain, $i^{th}$ spin is coupled to next $2\nu$ neighbours with
linearly decreasing exchange coupling. Note that the hamiltonian corresponding to $\nu=1$ is just the MG 
hamiltonian with dimer ground state. We will show that $H[{\mathfrak B}_{2\nu+1}]$, for any $\nu$, has the
 same ground state. Therefore, $\nu$ can be
 regarded as the label for the members in our family of 1D spin models with two-fold degenerate dimer ground state.

{\bf The ground state :} In order to find the ground state energy and the corresponding eigen-configurations of 
$H[{\mathfrak B}_{2\nu+1}]$, consider the 
minimum energy configurations for a ${\mathfrak B}_{2\nu+1}$ unit.
 Re-writing $h_i({\mathfrak B}_{2\nu+1})$ as 
\( \frac{1}{2}\left[\;\left(\S_i \right.\right.\) \(\left.\left. + \S_{i+1} + \cdots + \S_{i+2\nu}\right)^2\right. - \)
\( \left.\left(\S_i^2 + \S_{i+1}^2 + \cdots + \S_{i+2\nu}^2\right)\;\right] \)
tells us that the total spin, $\sum_{j=0}^{2\nu} \S_{i+j}$, being minimum corresponds to the lowest energy 
of the block. For $S=1/2$, the minimum energy is $-3\nu/4$, and the corresponding spin configurations are such that
 there is exactly one free spin while rest $2\nu$ spins forming a {\em block-singlet}.
A block-singlet can be described in terms of the bond-singlets (valence bond or dimers, as we often call them
so).
 Since there are many, independent ways of doing that, a block-singlet has an intrinsic degeneracy of
valence bond configurations. For example, ${\mathfrak B}_5$'s minimum energy configuration consists of a
 free spin, and a block-singlet of four spins. For $S=1/2$, there are exactly two linearly independent
valence bond configurations for a block-singlet of four spins. We will see later that in certain other models,
 this intrinsic degeneracy leads to a ground state with exponential degeneracy, and hence finite entropy density in 
the ground state. In the following, we try to construct the exact ground state configurations of 
$H[{\mathfrak B}_{2\nu+1}]$, for all values of $\nu$.

 If we can find {\em a configuration where every block of successive $(2\nu+1)$ spins share exactly one block-singlet
 of $2\nu$ spins, then it is the ground state configuration of} $H[{\mathfrak B}_{2\nu+1}]$. 
 This is ensured by the inequality, 
\begin{equation}
\left<\phi\right|H[{\mathfrak B}_{2\nu+1}]\left|\phi\right> \geq E_g \geq
\sum_{i=1}^Le_i^{min}({\mathfrak B}_{2\nu+1}) 
\label{inequality}
\end{equation}
 The following identities
 are used in establishing the fact that such a
construction will form an eigen-configuration of the $H[{\mathfrak B}_{2\nu+1}]$.
\begin{eqnarray}
\left(\S_{i_1} + \cdots + \S_{i_{2\nu}}\right)\cdot\S_{i_{2\nu+1}}\left[i_1,i_2,\dots,i_{2\nu}\right] &=& 0
                \label{ident1}\\
\sum_{j_1=1}^{2\nu-1}\sum_{j_2 > j_1}^{2\nu} \S_{i_{j_1}}\cdot\S_{i_{j_2}}\left[i_1,i_2,\dots,i_{2\nu}\right] &=& -\frac{3}{4}\nu \label{ident2}
\end{eqnarray}
In the Eq.(\ref{ident1}), \( i_{2\nu+1} \neq \left\{i_1,i_2,\dots,i_{2\nu}\right\} \).
The above identities are straight forward generalizations of what were used in finding
 the exact ground state for MG model for $S=1/2$. For general spin, $S$, the right hand side of 
Eq.(\ref{ident2}) will be $-\nu S(S+1)$. 
The notation \( \left[i_1,i_2,\dots,i_{2\nu}\right] \) denotes a block-singlet of $2\nu$ spins. Just to
illustrate this notation, consider four sites labeled as 1, 2, 3 and 4. Then [1,2,3,4] denotes all
the singlet configurations made up of spins sitting at these sites. Thus $[1,2,3,4]=[1,2]\otimes [3,4]$ or
$[2,3]\otimes [4,1]$. The indices
$\left\{i_1, i_2,\dots, i_{2\nu+1}\right\}$ take the values from set $\left\{i, i+1, \dots, i+2\nu\right\}$
while considering $i^{th}$ ${\mathfrak B}_{2\nu+1}$ block on a spin chain. Also, each one of the indices, 
$\left\{i_1, i_2,\dots, i_{2\nu+1}\right\}$, is distinct. Among the allowed dimer representations of 
\( \left[i_1,i_2,\dots,i_{2\nu}\right] \), one is simply \(
[i,i+1]\otimes[i+2,i+3]\otimes\cdots\otimes[i+2\nu-2,i+2\nu-1] \). This is like the dimers we have already seen.
There are many different types of them. Some will be of the type, say, 
\( [i,\;[i+1,i+2]\;,i+3]\otimes[i+4,i+5]\otimes\cdots\otimes[i+2\nu-2,i+2\nu-1] \).
Here, the notation \( [i,\;[i+1,i+2]\;,i+3] \) refers to a configuration where $\S_i$ and
$\S_{i+3}$ pair up to form a singlet while $\S_{i+1}$ and $\S_{i+2}$ doing the same. Each of these dimer
representations for 2$\nu$-spin block-singlet contains exactly $\nu$ dimers.
The fact that every site has identical exchange connectivity, suppresses the choices allowed by the intrinsic degeneracy
of block-singlets.
 The only configurations which satisfy the condition
 that every block of neighbouring $(2\nu+1)$ spins on a chain has exactly $\nu$ dimers, 
 are $\left|\psi_1\right>$ and $\left|\psi_2\right>$.
 In fact, the other configurations are not even the
eigen-configurations of $H[{\mathfrak B}_{2\nu+1}]$. One such configuration is shown in Fig.(\ref{dimers}-{\em c}).
This is so because such configurations always find some building blocks whose minimum energy
configuration is not satisfied, whereas $\left|\psi_1\right>$ and $\left|\psi_2\right>$ sastisfy all building blocks'
minimum energy configurations. Therefore, $\left|\psi_1\right>$ and
$\left|\psi_2\right>$ are the exact ground state configurations of the family of spin Hamiltonians given in
Eq.(\ref{hnu}), and the ground state energy is $-(3\nu J/4)L$.

 In the following, we give another proof of \( \left|\psi_1\right> \) and
\(\left|\psi_2\right>\) being the ground configurations of $H[{\mathfrak B}_{2\nu+1}]$
for all values of $\nu$, despite the fact that we have already shown it. This proof gives an independent
 existence to the hamiltonian $H[{\mathfrak B}_{2\nu+1}]$. 
The proof is based on the principle of
mathematical induction, and brings out an interesting property of {\em superstability}\cite{sutherland_shastry}
 which the dimer configurations possess. {\em An
eigen state, $\left|\phi\right>$, of some hamiltonian, $H$, is called superstable if it is also the eigen state
of the operator $H + V$, for a certain operator $V$ where the commutator, \( [H,V]\neq 0\)}. 
 To make a definitive statement about the superstability, we should clearly understand the relationship between 
successive hamiltonians of the family. The proof by induction is based on the understanding of such relationships
, and hence illustrates the superstability of the dimer configurations as the ground state of our 
family of hamiltonians, in a rigorous way.

{\bf The proof by induction} : For $\nu=1$, the hamiltonian operator is $H[{\mathfrak B}_3] = H_{MG}$. It is known that 
\(\HMG\left|\psi_{1,2}\right> = -(3J/4)L\left|\psi_{1,2}\right>\) with $\left|\psi_{1,2}\right>$ being the ground
state configuration. Here, $\left|\psi_{1,2}\right>$ refers to $\left|\psi_1\right>$ and $\left|\psi_2\right>$.
 Assume that $\left|\psi_{1,2}\right>$ is the eigen-configuration of $(\nu-1)^{th}$ member of the family,
 with eigen energy $E_{\nu-1}$ (for the sake of this proof ignore the fact that we have already proved it!). That is,
\(H[{\mathfrak B}_{2\nu-1}]\left|\psi_{1,2}\right> = E_{\nu-1}\left|\psi_{1,2}\right>\). Now, we check for the 
$\nu^{th}$ member. The hamiltonian for the $\nu^{th}$ member, where $\nu > 1$, can be re-written in the
following way.
\begin{eqnarray}
H[{\mathfrak B}_{2\nu+1}] & = & J\sum_{i=1}^L\sum_{j=1}^{2\nu}(2\nu+1-j)\S_i\cdot\S_{i+j}\nonumber\\
&=&H[{\mathfrak B}_{2\nu-1}] + J\sum_{i=1}^L\left\{\S_i\cdot\S_{i+2\nu} + 
                                                   2\sum_{j=1}^{2\nu-1}\S_i\cdot\S_{i+j}\right\}\nonumber\\
&=&H[{\mathfrak B}_{2\nu-1}] + \HMG + J\sum_{i=1}^L\left\{\S_i\cdot\S_{i+2} +
2\S_i\cdot(\S_{i+3}+\cdots+\S_{i+2\nu-1}) + \S_i\cdot\S_{i+2\nu}\right\}\nonumber\\
&=&H[{\mathfrak B}_{2\nu-1}] + \HMG + J\sum_{i=1}^L(\S_i + \S_{i+1})\cdot(\S_{i+3}+\cdots+\S_{i+2\nu})
\label{hnu_induction}
\end{eqnarray}
Clearly, from Eq.(\ref{hnu_induction}), \(H[{\mathfrak B}_{2\nu+1}]\left|\psi_{1,2}\right> = (E_{\nu-1}
-(3J/4)L)\left|\psi_{1,2}\right>\), as one can easily show that
\begin{equation}
\sum_{i=1}^L(\S_i + \S_{i+1})\cdot(\S_{i+3}+\cdots+\S_{i+2\nu})\left|\psi_{1,2}\right> = 0
\label{no_contribution_term}
\end{equation}
This proves that $\left|\psi_1\right>$ and $\left|\psi_2\right>$ are the eigen configurations of the
 $\nu^{th}$ member of the family,
with eigen energy, $E_{\nu} = E_{\nu-1} - (3J/4)L$ for any $\nu > 1$ ( with $E_1 = -(3J/4)L$).
 This gives $E_{\nu}=-(3\nu J/4)L$. Since $E_{\nu}$ we get here is same as the lower bound of the inequality
(Eq.(\ref{inequality})),
 the $\left|\psi_1\right>$ and $\left|\psi_2\right>$ are also the ground states, and not just the eigen states.
 Therefore, we have been able to prove that {\em the dimer configurations}, $\left|\psi_1\right>$ and
$\left|\psi_2\right>$, {\em form a superstable, two-fold degenerate ground subspace of the whole family of
$H[{\mathfrak B}_{2\nu+1}]$, parameterized by $\nu$, on a one dimensional lattice with PBC}.

The summary of our findings, in the context of one-dimensional spin systems, is the following:
\begin{itemize}
\item A family of spin models with exact dimer ground state is constructed. The ground subspace is two-fold
degenerate.
\item The general hamiltonian of the family is: \( H[{\mathfrak B}_{2\nu+1}] = \)
\( J\sum_{i=1}^L\sum_{j=1}^{2\nu}(2\nu+1-j)\S_i\cdot\S_{i+j} \). Here, $J>0$, $\nu$ is a positive integer and
${\mathfrak B}_{2\nu+1}$ refers to the fundamental building blocks for different members of the family.
Note that the hamiltonian has translational invariance, the exchange coupling decrease linearly with distance
between the coupled neighbours, and for a given $\nu$, every spin is coupled only upto $2\nu^{th}$ neighbour,
starting from the nearest one.
\item The ground state energy, in units of the strongest exchange coupling (the nearest neighbour coupling,
$2\nu J$), is $-\frac{3}{8}L$, for all members of the family.
\item The dimer states, $\left|\psi_1\right>$ and $\left|\psi_2\right>$, are superstable ground state
configurations with respect to all members of the family. 
\end{itemize}
Having described in detail the construction, and the exact ground state of this family of one dimensional 
spin models, let us briefly discuss the nature of elementary excitations of the same.

\begin{figure}
 \centerline{\epsfig{file=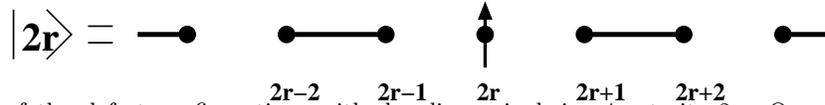,width=11cm}}
 \caption{This is one of the defect configurations with dangling spin being $\uparrow$, at site $2\mbox{r}$.
          One is allowed to have defect configurations with dangling spin polarization being $\downarrow$ without
          affecting the dispersion relation.}
 \label{kink} 
\end{figure}

{\bf The energy gap in the excitation spectrum :} The MG model has gap in the excitation spectrum with respect to 
the dimer ground state. This was illustrated by a variational calculation of the dispersion of the defect
 (the dangling spin as shown in Fig.(\ref{kink}) ) boundary between two exact ground state 
configurations\cite{ss_kink}. Later,  
it was exactly proved that there is an energy gap in the excitation above the exact dimer ground states of the MG
model\cite{AKLT_2}. A method to calculate the lower bound for the energy gap was also developed, and applied to
certain quantum antiferromagnets\cite{knabe}. 
We have calculated the dispersion relation for the propagating defect, variationally, for 
${\mathfrak B}_5$ and ${\mathfrak B}_7$ chains. We find that the energy gap towards such solitonic excitations 
exists for each of these members of the family, and seems to increase monotonically for higher members in the
 family. 

The dispersion relation for $H[{\mathfrak B}_3]$ (the MG model) is 
\( {\cal E}_3(k) = \frac{5}{4} + \mbox{Cos}(2k) \). This is already known from Ref.\cite{ss_kink}. 
Here, the lattice parameter, $a$, is taken to be unity, and $ -\pi/2 \leq k \leq \pi/2$.
 The single defect-boundary excitation energies for the next two members of the family, that is
$H[{\mathfrak B}_5]$ and $H[{\mathfrak B}_7]$, are the
following.
\begin{eqnarray}
{\cal E}_5(k) &=& 2 +\frac{3}{4}\mbox{Cos}(2k) -\frac{1}{2}\mbox{Cos}(4k)\label{dispersion_5} \\ 
{\cal E}_7(k) &=& \frac{11}{4} +\frac{3}{4}\mbox{Cos}(2k) -\frac{3}{8}\mbox{Cos}(4k) +\frac{1}{4}\mbox{Cos}(6k)
\label{dispersion_7}
\end{eqnarray}
Here, \({\cal E}_{2\nu+1} = \left<k\right|\left( H[{\mathfrak B}_{2\nu+1}] - E_{\nu} \right)
       \left|k\right>/\left<k\right|\left.k\right> \). $E_\nu = -\frac{3}{4}\nu N$; $N$ is the total number of ${\mathfrak
B}_{2\nu+1}$ units forming a finite chain of $L$ sites. $N=L$ for a chain with PBC, and $N=L-2\nu$ for an open chain.
Actually, we have considered the chains with odd number of spins, $L=4M+1$, in order to find the defect's
dispersion relation, and taken the limit $M\rightarrow\infty$.
 The ket, $\left|k\right>$
is defined as,
\begin{equation}
\left|k\right> = \frac{1}{\sqrt{2M+1}}\sum_{\mbox{r}=-M}^{M} \mbox{e}^{{\it i} k \mbox{r}} \left|2\mbox{r}\right>
\end{equation}
 The ket, $\left|2\mbox{r}\right>$, is a configuration where there is a dangling spin at 2r$^{th}$ site, and the rest
forming the dimer configuration of $\left|\psi_1\right>$ type on one side of the dangling spin and of
$\left|\psi_2\right>$ type on the other side, as shown in Fig.(\ref{kink}). The dangling spin can have either 
$\uparrow$ polarization or $\downarrow$ polarization. These defect configurations are non-orthogonal.
Since \( \left\{\left|2\mbox{r}\right>\right\} \) do not form a complete set of states, the propagating defect state, 
$\left|k\right>$, is only a variational choice. Nevertheless, it gives us some idea of the nature of excitations
above the ground state. From the dispersion relations for the defect, we find that the energy gap for ${\mathfrak B}_3$,
${\mathfrak B}_5$ and ${\mathfrak B}_7$ chains, in units of the nearest neighbour exchange, is 1/8, 3/16 and 11/48,
respectively. Presently, we are unable to identify any simple relation between the members of the family and
 the corresponding energy
gaps, nonetheless we see that there is a gap, and it seems to be increasing as we go up in the hierarchy.
 One can also consider the case where there are many such dangling spins in the dimer background, and consider the
 possibility of the bound states. We will not do it here. 

 {\bf The connection with the coulomb problem in 1D :} It is important to observe 
that the exchange coupling of the models constructed in the present work, 
$J_{ij} \propto (R - |i-j|)$, is exactly like the coulomb interaction in one dimension, albeit with a range $R$
\cite{acknowledge_shastry}.
 From $R^{th}$ neighbour onwards, the exchange coupling is zero. But there is no restriction on the range, $R$, 
and it can be
 anything. Hence, what we have found, essentially, is a quantum spin analogue of the coulomb problem in one dimension
 which was studied exactly by Lenard and Baxter long ago\cite{lieb_mattis}. It is an interesting and unexpected
connection. Analogous to the coulomb problem, one would expect plasmon like
gapped excitations in a spin model with infinitely long ranged, linear exchange coupling. 
For antiferromagnetic exchange coupling, as we saw just now, the energy gap increases with $R$. 
But the analogy has interesting consequences for the spin models with ferromagnetic linear exchange coupling.

 The hamiltonian for the ferromagnetic case is written as :
 \( H = -J\sum_{i=1}^{L}\sum_{m=1}^{R-1}(R-m)\S_i\cdot\S_{i+m} \). Here,
$J>0$, and $R$ can be even as well as odd unlike the antiferromagnetic case discussed in the present work.
 The ground state energy for a general 
spin S is, $E_g = -J S^2 R(R-1)L/2$. And the exact one magnon dispersion with respect to the ground state energy is : 
\( {\cal E}(k) = JS\left\{R^2 - [1-\mbox{Cos}(Rk)]/[1-\mbox{Cos}(k)]\right\} \), \( k = 2n\pi/L, \;\) where is $n$ is an
 integer. The wavenumber, $k$, takes values between $-\pi$ and $\pi$. 
For any finite $R$, ${\cal E}(k) \rightarrow 0$ quadratically, as $k\rightarrow 0$. Hence, the 
excitations are gapless. In order to consider the analogue of coulomb problem, 
$R$ should of the order of $L$, and let
$L$ go to infinity. We put $R=L/2$, and rescale $J$ to $J/R^2$ (as the ground state as well as the magnon
excitation energy goes as $R^2$ for large R). Then, \( {\cal E}(k) = JS\left\{1-\frac{4[1-(-)^n]}{L^2[1-Cos(k)]}
\right\} \).
Therefore, in the thermodynamic limit, the magnon excitation has a gap of value $JS$, and a totally flat dispersion.
Hence, a ferromagnet with infinite ranged linear exchange coupling is gapped. Next, we discuss another type of
1D spin models where the dimer ground state is exponentially degenerate. These models are also constructed of
${\mathfrak B}_{2\nu+1}$ units.

{\bf 1D models with exponentially degenerate dimer ground state :} As mentioned earlier, the block-singlets made
 up of four or more spins always have degenerate dimer representations. This intrinsic
degeneracy at the block level, however, could not be greatly exploited in our previous construction which led to the
class of linear exchange spin models on a chain. The dimer ground state of this class of models is
only two-fold degenerate. Now, we construct another type of spin models, on a closed chain, whose ground state
has exponentially large number of degenerate dimer configurations. Here, the degeneracy of ground state is
exponential in the number of lattice sites, $L$. To illustrate this class of models, we describe a particular 
construction using ${\mathfrak B}_5$ units.
\begin{figure}
 \centerline{\epsfig{file=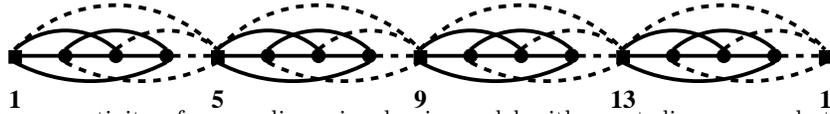,width=11cm}}
 \caption{The exchange connectivity of a one dimensional spin model with exact dimer ground state whose
          degeneracy is exponential in the number of sites. Shown in the figure is an example of 16 site model
          on a closed chain geometry. The sites where ${\mathfrak B}_5$ unit repeats itself are numbered explicitly.
          The exchange couplings are : solid line $\equiv \alpha J$ and dashed line $\equiv J$.}
 \label{1dmodel}
\end{figure}
Consider a closed chain with even number of sites. Connect spins at 1$^{st}$, 2$^{nd}$, 3$^{rd}$ and 4$^{th}$ 
sites identically among themselves with exchange coupling $\alpha J$. Then connect each of these four spins to the 
spin at 5$^{th}$ site with exchange coupling $J$, as shown in Fig.(\ref{1dmodel}).
 Again, connect spins from 5$^{th}$ to 8$^{th}$ site identically with 
coupling $\alpha J$, and then connect these four spins to the spin at 9$^{th}$ site with coupling $J$. Repeat
this procedure for further spins, starting with 13$^{th}$ site, 17$^{th}$ and so on. Thus we construct a spin
chain of ${\mathfrak B}_5$ units. These ${\mathfrak B}_5$ units are slightly different from the ones we have used
earlier. Here we have two exchange couplings $J$ and $\alpha J$ unlike the earlier considerations 
where coupling was identically $J$. The corresponding
hamiltonian can be written as : \(H = J\sum_{l=1}^{L/4}h^{\prime}_{4l-3}({\mathfrak B}_5)\), with $J>0$. The block
hamiltonian, 
\begin{eqnarray}
h^{\prime}_{4l-3}({\mathfrak B}_5) & = & \alpha \{\S_{4l-3}\cdot (\S_{4l-2} + \S_{4l-1} + \S_{4l}) +
                     \S_{4l-2}\cdot (\S_{4l-1} + \S_{4l}) + \S_{4l-1}\cdot\S_{4l} \} \nonumber\\
& & + \S_{4l+1}\cdot (\S_{4l-3} + \S_{4l-2} + \S_{4l-1} + \S_{4l})\label{hprimeb5}
\end{eqnarray}
is a slightly generalized version of the previously defined ${\mathfrak B}_5$ block hamiltonian, $h({\mathfrak
B}_5)$. When $\alpha = 1$, \( h^{\prime}({\mathfrak B}_5) \) is same as \( h({\mathfrak B}_5) \). 

For $\alpha > 1$, \( h^{\prime}_{4l-3}({\mathfrak B}_5) \)'s minimum energy is equal to $-3\alpha/2$ for $S = 1/2$. 
Spins, $\S_{4l-3}$, $\S_{4l-2}$, $\S_{4l-1}$ and $\S_{4l}$ form a two-fold degenerate block-singlet
corresponding to block hamiltonian's lowest eigen energy, and spin $\S_{4l+1}$ remains ``free''
 to be a part of the next block-singlet. Therefore, the ground state energy of this spin model
is $-3\alpha JL/8$, and the ground state configurations can be written as:
\begin{equation}
\left|\psi_g\right> = [1,2,3,4]\otimes[5,6,7,8]\otimes\cdots\otimes[L-3,L-2,L-1,L]
\label{b5_gnd1}
\end{equation}
Since each of these four spin block-singlet, \(\left[4l-3,4l-2,4l-1,4l\right]\), has two dimer representations, 
$\left|\psi_g\right>$ represents $2^{L/4}$ degenerate, dimer configurations forming the ground subspace. For
$\alpha = 1$, the ground state energy is still given by the above expression, but the degeneracy of the ground
state is 2(2$^{L/4}$). This is so because, for $\alpha=1$, any four spins of a ${\mathfrak B}_5$ block can form
singlet in the minimum energy configuration. This allows following configurations, together with \(
\left|\psi_g\right> \), in the ground state.
\begin{equation}
\left|\psi_g^\prime\right> = [2,3,4,5]\otimes[6,7,8,9]\otimes\cdots\otimes[L-2,L-1,L,1]
\label{b5_gnd2}
\end{equation}
And hence, the degeneracy is doubled. 
The procedure, described here in detail, can be directly applied to construct models using bigger spin blocks. 
Thus we are able to construct quantum spin chains with exponentially degenerate dimer ground state, and
interesting thing to note is that the ground state is exactly known for $\alpha \ge 1$, unlike the linear
exchange models where $\alpha$ is strictly one. Or, in other words, the dimer configurations are superstable for
all values of $\alpha$ greater than or equal to one. 
We will come across the same features once more in the following section, while considering the 
higher dimensional generalization of our scheme. 

In this section, we have developed the concepts, and explicitly used them in constructing one dimensional 
spin models. In the next section, we will use this understanding to construct models with exact dimer ground state
 in two and three dimensions. Before going to the next
section, we must mention that recently there have been some generalizations of MG and SS models
\cite{shankar,buttner,balents}. In this regard, we would like
to stress upon the point that our scheme provides a very general framework to construct all such models using
only two things : (1) the basic building blocks, and (2) arranging them in such a way that each block is able
to satisfy its minimum energy configuration even when being a part of the full assembly.
This we have already seen working in one dimension.  We will illustrate this in the following section by 
constructing various models in two and three dimensions.
\section{The generalization to two and three dimensions}
The SS model is a two dimensional spin model with a specific exchange connectivity\cite{shastry_sutherland}.
 It has an exact dimer ground state with no degeneracy. It is, in some sense, the two dimensional 
analogue of the MG model. 
 The SS model, like the MG model, is made up of fundamental blocks of three spins, that is, 
${\mathfrak B}_3$ units in our language. Hence, SS model is the first member
of the family of 2D spin models with an exact dimer ground state, if labeled according to our scheme. 

This correspondence motivates us to construct spin models in higher dimensions
 whose ground states can be known exactly. In fact, we can construct a whole lot of them. 
The rules are very simple. 
{\em Pick one spin of a ${\mathfrak B}_{2\nu+1}$ unit as free. Leave rest to form a block-singlet of size $2\nu$.
Each of the $2\nu$ spins of the singlet forming block can act as the ``free" spin for the neighbouring blocks.
And any given block-singlet can be fully or partially shared by other ``free" spins}.
 This allows us to extend the network in higher dimensions. The hamiltonian for any such model is just the sum of
 all the block-hamiltonians. {\em Only those dimer configurations which satisfy each building block's minimum
energy configuration form the exact ground state configurations.}
This set of rules seem sufficiently general for constructing models with exact dimer ground states. For example, one
gets the SS model by applying these rules to make 2D model using ${\mathfrak B}_3$ units.

\begin{figure}
 \centerline{\epsfig{file=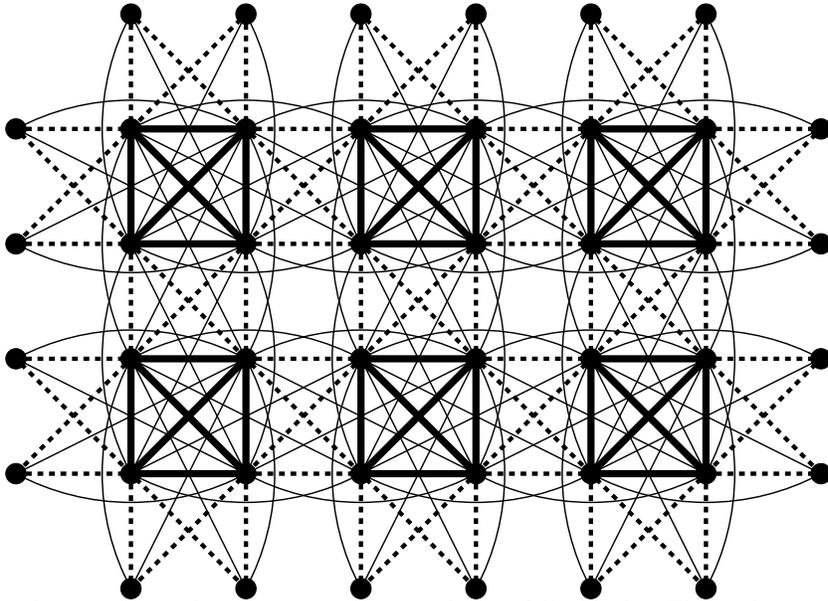, width=11cm}}
 \caption{A 2D model with a certain exchange connectivity, made up of ${\mathfrak B}_5$ blocks. The exchange
          couplings are : thin solid line $\equiv J$, dashed line $\equiv 2J$, and thick solid line
          $\equiv 8\alpha J$.}
 \label{2d_model}
\end{figure}

\begin{figure}
 \centerline{\epsfig{file=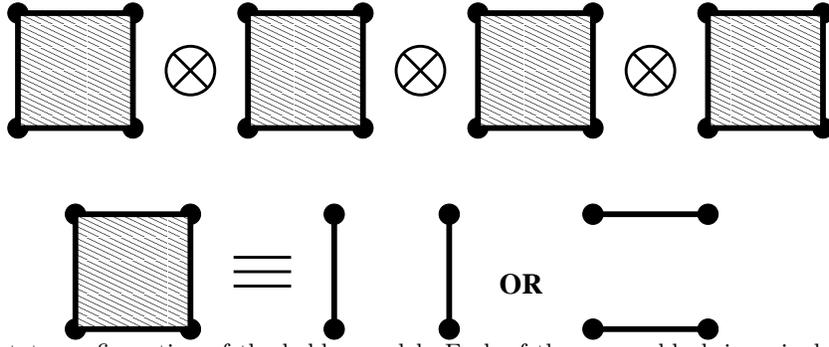, width=11cm}}
 \caption{The ground state configuration of the ladder model. Each of the square block is a singlet which has
          two linearly independent dimer representations as shown by vertical and horizontal dimers.}
 \label{ladder_ground_state}
\end{figure}

We consider a generalized ${\mathfrak B}_{2\nu+1}$ unit, where one of the spins is picked up to be exclusive
such that it is coupled to the rest by unit exchange strength while the rest are completely coupled among
 themselves by some exchange $\alpha$. These basic spin blocks are still completely connected, 
but the couplings are not identical. This choice is exactly like the one we had for constructing 1D models with 
exponentially degenerate dimer ground state. 
See Fig.(\ref{B3_B5}), where ${\mathfrak B}_3$ and ${\mathfrak B}_5$ units are shown in one, two and
three spatial dimensions.
The ``block-singlet + free spin" is the lowest energy configuration of the ${\mathfrak B}_{2\nu+1}$ unit for
$\alpha \geq 1$. We can construct models in 2D (and also in 3D) in such a way that {\em spatially disjoint 
block-singlets} form the exact ground state configurations for $\alpha \geq 1$.  
It is interesting to note that the domain of superstability for the dimer ground states in these models is
 $\alpha\geq 1$, unlike the case of translationally invariant 1D models constructed in the previous section.
 There, the dimer configurations form a superstable ground state only at $\alpha=1$.
Also, the spatial disjointness preserves the intrinsic degeneracy of 
the block-singlets which makes the ground state exponentially degenerate. Thus, we are able to
construct {\em spin models with finite entropy density in the ground state}.
 To illustrate all this, we describe models made up of ${\mathfrak B}_5$ units. 
\begin{figure}
 \centerline{\epsfig{file=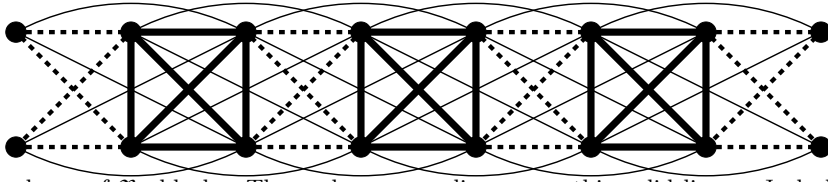, width=11cm}}
 \caption{The ladder made up of ${\mathfrak B}_5$ blocks. The exchange couplings are : thin solid line $\equiv
          J$, dashed line $\equiv 2J$, and thick solid line $\equiv 4\alpha J$.}
 \label{ladder}
\end{figure}
Following the rules stated above, a ladder model, as shown in Fig.(\ref{ladder}), is constructed with 
${\mathfrak B}_5$ units (we could have constructed a simpler ladder with ${\mathfrak B}_3$ units, but the
exponentially degenerate ground state can not be realized there. The dimers along the rungs form the
exact ground state configuration of a ${\mathfrak B}_3$ ladder.).
These ladders properly arranged on a plane, give rise to a certain 2D spin model as shown in Fig.(\ref{2d_model}). 
Assuming a ribbon like geometry for the ladder and the toroidal geometry for the 2D model, we can easily 
find their ground state energies. In our construction, each lattice point of the ladder model contributes
exactly one ${\mathfrak B}_5$ unit, whereas each site on the 2D model contributes two such units. Therefore, the
ground state energies, for $\alpha\geq 1$, of the ladder and the 2D model are $-\frac{3}{2}\alpha J L$ and 
$-3\alpha J L$, respectively.
 Here, $L$ is the total number of sites. The ground state energy per site, in units of the 
strongest exchange
($4\alpha J$ for ladder, and $8\alpha J$ for 2D model), is just $-\frac{3}{8}$. 
The blocks of spins, connected with thick lines as shown in the Figs. (\ref{ladder}) and (\ref{2d_model}),
 forming singlets is the
exact ground state configuration. For the ladder, the ground state configuration is shown in
Fig.(\ref{ladder_ground_state}). The ground state for the 2D model is similar to that of the ladder. 

The two-fold intrinsic degeneracy of each block-singlet makes the 
ground subspace $2^{L/4}$-fold degenerate. The entropy density in the ground state is $\frac{1}{4}
\mbox{log}(2)$, just one fourth of that of a paramagnet. Each block-singlet has two independent 
bond-singlet configurations, therefore, the ground subspace consists of $2^{L/4}$ distinct dimer 
 or valence bond configurations. For example, two of these configurations are the 
columnar dimer states which are the exact ground states of a certain model constructed by Bose and
Mitra\cite{bose}. Thus, the ground state of models discussed here is a spin-liquid as well as a dimer-liquid,
 as the dimers within a block-singlet are correlated, but there is no correlation between dimers belonging to
 different block-singlets. One can easily construct certain other models, using same rules, whose ground states
are dimer solids, though we will not discuss these models explicitly.

\begin{figure}
 \centerline{\epsfig{file=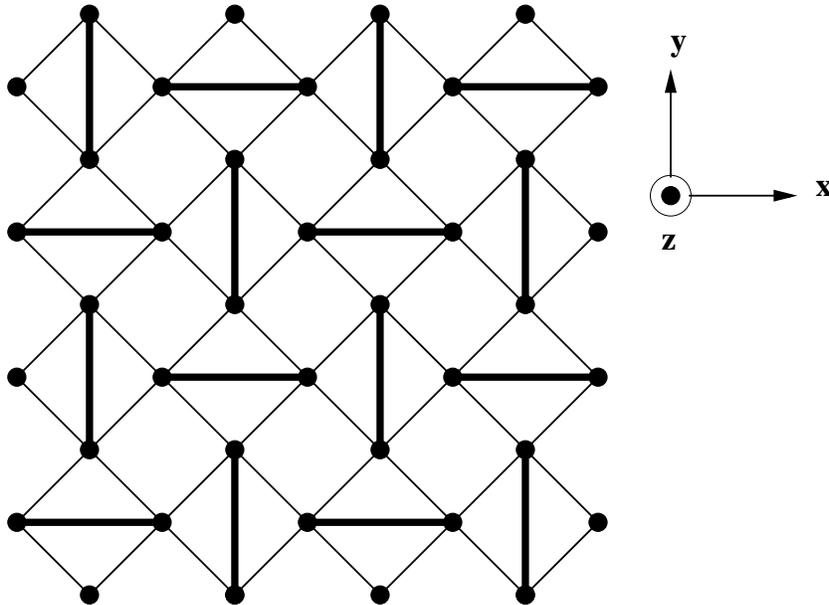,width=11cm}}
 \caption{This is a projection on x-y plane, of the ground state configuration of a layer of the corner sharing
           octahedra made out of square pyramids (the ${\mathfrak B}_5$ units in 3D). Thick lines, here, represent
           the common basal plane of two ${\mathfrak B}_5$ pyramids making an octahedra.
           In the ground state, four spins lying on the common basal plane form a two-fold degenerate
           block-singlet. These block-singlets lie in x-z or y-z planes. It is interesting to observe the
           topological equivalence of the arrangement of block-singlets on orthogonal planes, here, to that of the
           dimers in the exact ground state of the SS model.}
 \label{3d_model}
\end{figure}

 The generalizations of SS model, considered
in \cite{shankar}, provide good examples of the higher dimensional models made up of ${\mathfrak B}_3$ units.
So, we will consider the next block, that is, ${\mathfrak B}_5$. 
The three dimensional (3D) ${\mathfrak B}_5$ block is
shown in Fig.(\ref{B3_B5}). There can be many ways of constructing 3D models using 
${\mathfrak B}_5$-pyramids. We consider a case where two such pyramids share the basal plane to form an
octahedra. For $\alpha \geq 1$, the four spins of the common basal plane form a block-singlet.
 Layers of the corner-sharing octahedra, stacked in certain ways, form one such model. The projection of one
such layer on x-y plane is shown in Fig.(\ref{3d_model}). The ground state configurations are such that
 the singlet blocks are lying in orthogonal planes. 
It is topologically similar to the orthogonal arrangement of dimers in the exact ground state of 
2D SS model except that these are block-singlets, and are lying in orthogonal planes.
 The spatial disjointness of the block-singlets belonging to different planes gives 
rise to exponentially degenerate dimer ground state.

The possibilities for making such models are enormous, and therefore can not be exhausted in one place.
 But one thing must be emphasized upon. The basis for constructing models with exact dimer ground states, as
discussed in the present work, is very general.
 The model hamiltonians that may be constructed with these rules, may be of definite
importance in understanding some physics of the spin systems.
\section{Conclusion}
We, first, summarize the main results. We have constructed a family of one-dimensional spin models with linearly
decreasing exchange coupling. The range of coupling is the family parameter which is decided by the
size of the basic building block. The fundamental building blocks are completely connected blocks of odd number of
identical spins. All the members in the family have an exact, two-fold degenerate dimer ground state.
There exists an energy gap in the excitation spectrum above the dimer ground states, and it seems to be increasing
monotonically for the higher members in the family. The interesting analogy of this family of models, with the 
coulomb problem in one dimension is discussed. One dimensional spin models with exponentially degenerate 
dimer ground states are also discussed.

 This way of constructing models is easily generalized to higher dimensions.
 We are able to construct models in two and three dimensions, with exact dimer ground states. 
The ground state of many such models is exponentially degenerate. The degeneracy in the ground
state of these models arises due to many degenerate dimer representations of the
block-singlets forming the ground state configuration. These models, thus, provide explicit examples of the  
systems with finite entropy density in the ground state, and hence violating the third law of thermodynamics. Also,
these are the models with {\em finite ranged }RVB type ground states, spanned by an exponentially large subset of 
the full set of linearly independent valence-bond configurations. 
The range of the valence-bonds is decided by the size of the
building block. The rules for constructing such models are explicitly discussed.

We hope that the general class of models proposed here may be useful in studying the quantum phase
transitions in the frustrated spin systems. The exact knowledge of the ground state with exponential degeneracy,
in these models, is particularly remarkable. The building block way of looking at these models is quite in the
spirit of synthetic chemists. It may be possible to realize such models.
\section{Acknowledgement}
The author wishes to thank Professors B. Sriram Shastry and Diptiman Sen for fruitful discussions 
and comments, and likes to acknowledge the Council of Scientific and Industrial Research (CSIR), India, for financial support.

\end{document}